\documentclass[runningheads]{llncs}
\usepackage[T1]{fontenc}
\usepackage{graphicx}
\usepackage{latexsym}
\usepackage{amssymb}
\usepackage{amsmath}
\usepackage{booktabs}
\usepackage{enumitem}
\usepackage{graphicx}
\usepackage{tabularx}
\usepackage{color}
\usepackage{tikz}
\usetikzlibrary{shapes}

\usepackage{pifont}
\newcommand{\cmark}{\ding{51}}%
\newcommand{\xmark}{\ding{55}}%

\begin{document}
\title{Parameterized Voter Relevance \\ in
Facility Location Games  \\ with Tree-Shaped Invitation Graphs}
\titlerunning{Parameterized Voter Relevance in Facility Location with Invitation}
%
\author{Ryoto Ando\inst{1} \and
Kei Kimrua\inst{1}
\and
Taiki Todo\inst{1}
\and
Makoto Yokoo\inst{1}
}
%
\authorrunning{}
%
\institute{
Graduate School of ISEE, Kyushu University\\
Motooka, Fukuoka 819-0395 Japan 
}
%
\maketitle              
\begin{abstract}
 Diffusion mechanism design, which investigate how to incentivise agents to invite as many colleagues to a multi-agent decision making as possible, is a new research paradigm at the intersection between microeconomics and computer science.
 In this paper we extend traditional facility location games into the model of diffusion mechanism design. Our objective is to completely understand to what extent of anonymity/voter-relevance we can achieve, along with strategy-proofness and Pareto efficiency when voters strategically invite collegues. 
 We define a series of anonymity properties 
 applicable to the diffusion
 mechanism design model, as well as parameterized voter-relevance properties for guaranteeing reasonably-fair decision making.
 We obtained two impossibility theorems and two existence theorems, which partially answer the question we have raised 
 in the beginning of the paper.

\keywords{Mechanism Design  \and Facility Location Games \and Information Diffusion}
\end{abstract}
%
%
%
\section{Introduction}
\label{sec:intro}

Social choice theory is one of the mathematical foundations of 
multi-agent decision making. We assume there exists a number of individuals, usually called as {\em agents} or {\em voters}.
Each voter has a preference over a set of alternatives/outcomes,
and a social choice function chooses, by taking into account the preferences of voters, an 
alternative 
as a final outcome. As many impossibility theorems have been obtained in the 
literature~\cite{satterthwaite:strategyproof:85}, designing social choice functions for making an appropriate social decision has been an
important research topic. 

Facility location games are a well-studied problem in the literature of
social choice and known as a special case of voting~\cite{moulin:PC:1980}.
In the problem, each agent (voter) is located at a point on an interval that represents the set of social alternatives.
Under the realization of a social alternative 
as the outcome of a social choice function, 
a voter's cost is defined as the distance between the outcome and her location.
The domain restriction on such {\em single-peaked} preferences guarantees the existence 
of a Condorcet winner. Actually the alternative 
most preferred by the {\em median voter}, i.e., the voter whose 
location is the $\lfloor (n+1)/2 \rfloor$-th smallest
among $n$ voters, is a Condorcet winner.

In the literature of mechanism design,
truthfulness, also known as {\em strategy-proofness}, is 
one of the most important properties
that a social choice function should preserve.
It requires that
telling a preference honestly to the social choice function
is a dominant strategy for every voter.
Clarifying the necessary and sufficient condition
for a social choice function to be strategy-proof has
greatly attracted considerable attention of researchers.
The {\em median voter schemes}, which contains the above median rule 
as a special case, are the only social choice functions 
that satisfy strategy-proofness, ontoness, and anonymity~\cite{moulin:PC:1980}.

In practice, 
a mechanism designer, which we also call a {\em moderator},
struggles to completely observe the set of voters who are 
willing to participate in the decision making process and to directly advertise it to them. 
Instead, 
voters will usually get the information about a decision making process
from their 
peers/friends, through their shared {\em social networks}, sometimes also called as {\em invitation graphs}.
It is therefore important to incentivize voters to
invite as many colleagues as possible.
Such a new paradigm 
of mechanism design is called {\em diffusion mechanism design}~\cite{li:AAAI:2017}. 
However, as far as the authors know,
any literature on social choice theory, including existing works on facility location games,
have never considered diffusion mechanism design.

According to these situations, 
in this paper we consider social choice under single-peaked preferences over networks. The main goal in this research direction is to completely understand which social choice functions satisfy/do not satisfy strategy-proofness and Pareto efficiency, as well as some anonymity properties. 
As a first step, we define a class of anonymity properties applicable
to the model of mechanism design via social networks.
Specifically, we define three anonymity properties, namely anonymity on structures (AN-S), anonymity on distances (AN-D), and 
anonymity on structure-distance pairs (AN-SD), each of which considers
the structure of social networks in their own definitions.

To understand whether 
socially-fair decisions can be made, 
we also define a parameterized class of {\em voter-relevance}
properties and discuss which combinations between anonymity and voter-relevance are compatible with SP and PE.
A social choice function is $d$-distance voter-relevant
(VR-$d$ in short) 
if any voter who is at distance $d \in \mathbb{N}_{\geq 0}$ or less from the moderator
has {\em some chance} to change the final outcome by changing her own action.
Obviously, achieving $d = n$ is the best when $n$ voters participate; any voter is at distance $d$ or less then, i.e.,
all the voters have some chance to change the outcome.
We obtained two impossibility and two existence theorems
when 
the social networks among voters are
tree-shaped. 
Especially, these results give us a complete understanding on to what extent of voter-relevance we can achieve when AN-S/AN-D are set as mandatory, as shown in the two center columns of Table~\ref{tbl:summary}.

The rest of this paper is organized as follows.
Subsection~\ref{sec:literature} reviews the related literature
of social choice theory and diffusion mechanism design.
Section~\ref{sec:model} defines the standard model
of facility location games and its extension with
strategic information diffusion.
Section~\ref{sec:ANS} shows an impossibility theorem
on anonymity on structures (AN-S), corresponding to the 
AN-S column of the table.
Section~\ref{sec:AND} shows an impossibility theorem
and an existence theorem on anonymity on distances (AN-D),
corresponding to the AN-D column.
Section~\ref{sec:ANSD} shows an existence theorem
on anonymity on structure-distance pairs (AN-SD),
corresponding to the AN-SD column.
Section~\ref{sec:discussion} concludes the paper.



\begin{table}[tb]
 \centering
 \caption{Summary of existence of SCFs satisfying strategy-proofness over social networks and Pareto efficiency, along with corresponding anonymity and voter-relevance properties. 
 Requirements basically get weaker by moving from the top-left corner
 to the bottom-right corner, while
 there is one exception that there is no implication relations between
 the two center columns, AN-S and AN-D.
 }
 \label{tbl:summary}
 \def\arraystretch{2}
 \begin{tabular}{c|c|c|c|c}
 ~~~~~~~~~~~~~~~~~~~    & ~~~~ AN ~~~~    & AN-S & AN-D & AN-SD \\ \hline
 VR-n & \xmark & \xmark & \xmark & Open \\ \hline
 VR-3 ... VR-n-1 & \xmark & \xmark & \xmark & Open \\ \hline
 VR-2 & \xmark & \xmark & ~ \xmark \ (Thm~\ref{thm:imp:AND-R2}) ~ & ~ \cmark \ (Thm~\ref{thm:p:ANSD-R2}) ~ \\ \hline
 VR-1 & \xmark & \xmark & ~ \cmark \ (Thm~\ref{thm:p:AND-R1}) ~ & \cmark \\ \hline
 VR-0 & \xmark & ~ \xmark \ (Thm~\ref{thm:imp:ANS-R0}) ~ & \cmark & \cmark \\ \hline
 \end{tabular}
\end{table}

\section{Literature}
\label{sec:literature}

Under the single-peaked preferences, Moulin~\cite{moulin:PC:1980} investigated strategy-proof and Pareto efficient social choice functions on a continuous line and proposed a class of such social choice functions, so-called generalized median voter schemes.
Indeed, it is the only class of deterministic,
strategy-proof, Pareto efficient and anonymous social choice functions.
Procaccia and Tennenholtz~\cite{procaccia:TEAC:2013}
initiated the research 
on approximation mechanism design, for which a facility location problem was chosen as a case study. 
Recently, some new models of facility locations have been 
investigated, including 
dynamic facility locations~\cite{KeijzerW:IJCAI:2018,wada:AAMAS:2018}
and locating multiple 
facilities~\cite{serafino:ECAI:2014,fong:AAAI:2018,anastasiadis:AAMAS:2018}.
Some other research 
also considered the strategy-proof facility location on discrete structures, such as grids~\cite{sui:IJCAI:2013,escoffier:ADT:2011}
and cycles~\cite{alon:MOR:2010,alon:DM:2010,dokow:EC:2012}.

The research of diffusion mechanism design, also known as {\em mechanism design over social networks},
was initiated by 
Li et al.~\cite{li:AAAI:2017},
which
considered single-item auctions and proposed 
a strategy-proof mechanism.
After that, 
several works investigated 
strategy-proof
resource allocation mechanisms with monetary compensations, e.g., multi-unit auctions and redistributions~\cite{zhao:AAMAS:2018,kawasaki:AAAI:2020,li:IJCAI:2020,zhang:AAMAS:2020,JEONG2024191}.
On the other hand, there is limited research on 
decision making without money from the perspective of diffusion mechanism design.
Some recent works have house allocation problems without monetary compensation~\cite{kawasaki:AAMAS:2021,you:AAMAS:2022}
and two-sided matching (also known as school choice) problem~\cite{ChoTY:IJCAI:2022}.
However, 
neither papers have
addressed voting/social choice functions 
from the perspective of diffusion mechanism design.

\section{Model}
\label{sec:model}

We begin with defining the traditional model of social choice with single-peaked preferences, which is followed by the extended model with information diffusion.


There is a line segment $\mathcal{O} := [0,1]$, from which a single point/outcome should be chosen by a social choice function.
Since in our model the set of participating voters depends on
the information diffusion by voters, 
we need to define both {\em potential} voters and {\em participating} voters.
Let $\mathcal{N}$ be the set of potential voters,
and let $N \subseteq \mathcal{N}$ be a set of participating voters.
Each voter $i \in N$ has a {\em preference} $\succ_{i} \in \mathcal{P}$, which is represented as a complete and transitive binary relation over $\mathcal{O}$.
Let $P$ represent the set of possible preferences over $\mathcal{O}$.
In this paper we assume preferences are {\em single-peaked}, which is one of the well-studied preference restriction in the literature.

\begin{definition}[Single-Peaked Preferences]
 \label{def:single-peaked}
 A domain $\mathcal{R} \subseteq \mathcal{P}$ of preferences over 
 the set of (possibily infinite) alternatives $\mathcal{O}$ is 
 {\em single-peaked} if there is a strict ordering $\rhd$ over $\mathcal{O}$ s.t.\ 
 for any preference $\succ \in \mathcal{R}$, there is an associated ideal alternative $p \in \mathcal{O}$,
 and for any two distinct alternatives $x,y \in \mathcal{O}$, 
 $x \succ y$ if and only if either $p \rhd x \rhd y$ or $y \rhd x \rhd p$ holds.
\end{definition} 

That is, 
a voter who has preference $\succ_{i}$ associated with peak $p_{i}$ prefers 
outcome $x$, which is strictly closer to (resp.\ farther from) $p_{i}$
than another outcome $y$.
When voter $i$ has preference $\succ_{i}$
associated with the peak $p_{i} \in \mathcal{O}$,
we sometimes say that voter $i$ is {\em located at} $p_{i}$.
Let $\succ := (\succ_{i})_{i \in N}$
denote a profile of the voters' preferences, and 
let $\succ_{-i} := (\succ_{i'})_{i' \neq i}$ denote 
the profile without $i$'s.

A (deterministic) social choice function is 
a mapping from the set of possible profiles to the set of outcomes.
Since the number of participating voters
varies with regard to the voters' actions,
a social choice function must be defined for different-sized 
profiles. To describe this feature, 
we define a social choice function $f = ({N})_{N \subseteq \mathcal{N}}$ 
as a family of functions, where each $f^{N}$ is a mapping from $\mathcal{R}^{|N|}$ to
$\mathcal{O}$.
When a set $N$ of voters participates,
the social choice function $f$ uses function $f_{N}$ to
determine the outcome.
The function $f_{N}$ takes profile $\succ$ of preferences
jointly reported by $N$ as an input, 
and returns $f^{N}(\succ)$ as an outcome.
We denote $f^{N}$ as $f$ if it is clear from the context.

\begin{definition}[Social Choice Function (SCF)]
 \label{def:scf}
 A {\em social choice function} (SCF) $f = (f^{N})_{N \subseteq \mathcal{N}}$ is 
 a family of mappings $f^{N}$ from $\mathcal{R}^{n}$ to $\mathcal{O}$, where $|N| = n$,
 s.t.\ each $f^{N}$ takes $n$ preferences as an input and returns an alternative $o \in \mathcal{O}$.
\end{definition}

Strategy-proofness, which is one of the most important properties that SCFs should satisfy, requires that for each voter,
reporting her true preference to the social choice function is a dominant strategy, i.e., one of the best action regardless of the action profile of
other voters. 

\begin{definition}[Strategy-Proofness]
 \label{def:sp}
 An SCF 
 $f$ is 
  {\em strategy-proof} if,
 for any $N \subseteq \mathcal{N}$ s.t.\ $|N| = n$,
 any $i \in N$,
 any $\succ'_{-i}$,
 any $\succ_{i}$,
 and any $\succ'_{i}$,
 $f(\succ_{i}, \succ'_{-i})
 \succsim_{i}
 f(\succ'_{i}, \succ'_{-i})$.
\end{definition}

Ontoness is a minimum requirement to guarantee that the decision making process is fair for outcomes, which requires that for any outcome,
there is at least one profile of preferences
under which it is chosen.
If ontoness is not satified, i.e., there is an outcome that cannot be chosen under any profile,
then it is reasonable to remove it from the set of outcomes.

\begin{definition}[Ontoness]
 \label{def:onto}
 An SCF $f$ is 
 {\em onto} if,
 for any $o \in \mathcal{O}$,
 there is at least one profile $\succ$ s.t.\ 
 $f(\succ) = o$. 
\end{definition}

Anonymity on preferences, traditionally just called as anonymity, is, on the other hand,
a requirement on fairness among voters.
Under an SCF that is anonymous
on preferences, a permutation of voters' names/identities does not affect the chosen outcome. 

\begin{definition}[Anonymity]
 \label{def:anon}
 An SCF $f$ is 
 {\em anonymous} if,
 for any profile $\succ$ and any permutation $\succ'$ of $\succ$,
 $f(\succ) = f(\succ')$ holds.
\end{definition}

Moulin proposed a class of SCFs satisfying strategy-proofness and ontoness.
Furthermore, he showed that any SCF that satisfies both of these properties
can be represented as an instance in this class.

\begin{definition}[Generalized Median Voter Schemes (GMVS)~\cite{moulin:PC:1980}]
 \label{def:gmvs}
 A {\em generalized median voter scheme (GMVS)} $f$ is 
 an SCF defined as follows:
 $\forall N \subseteq \mathcal{N}$ s.t., $|N| = n$,
 there is a profile 
 $\alpha^{N} = (\alpha^{N}_{S})_{S \subseteq N} \in \mathcal{O}^{2^{n}}$
 of $2^{n}$ parameters
 s.t.\ 
 \begin{enumerate}
  \item $\alpha^{N}_{\emptyset} = 0$, $\alpha^{N}_{N} = 1$
  \item $\alpha^{N}_{S} \leq \alpha^{N}_{T}$ for any $S \subseteq T \subseteq N$, and
  \item $f^{N}(\succ) = \max_{S \subseteq N} \min \{(p_{i})_{i \in S}, \alpha^{N}_{S} \}$ for any input $\succ \in \mathcal{R}^{n}$.
 \end{enumerate}
\end{definition}

\begin{theorem}[Moulin~\cite{moulin:PC:1980}]
 \label{thm:moulin}
 Under the single-peaked preference domain, an SCF satisfies 
 strategy-proofness and ontoness
 if and only if it is a generalized median voter scheme.
\end{theorem}

\subsection{Mechanism Design via Social Network}
\label{ssec:diffusion}

In our model of facility location games, 
voters are distributed over a social network (or an {\em invitation graph}),
which is assumed to be a directed tree with a single source vertex.
There is a special agent called {\em moderator},
represented as symbol $m$ and corresponds to the source vertex.
Let $r_{m} \subseteq \mathcal{N}$ be the set of the children of moderator $m$, 
which are also called the {\em direct children} of $m$.
For each voter $i \in \mathcal{N}$, let $r_{i} \subseteq \mathcal{N} \setminus \{i\}$
denote $i$'s children. 
Given $r_{\mathcal{N}} := (r_{i})_{i \in \mathcal{N}}$ and $r_{m}$, all the parent-child relations are defined, 
specifying the {\em social network} $G(r_{\mathcal{N}}, r_{m})$ among voters and the moderator.

Next we
give some additional notations 
to formalize our model as a mechanism design problem.
Let $\theta_{i} = (\succ_{i}, r_{si})$ denote the {\em true type} of voter $i$, and 
let $\theta = (\theta_{i})_{i \in \mathcal{N}}$ denote a type profile of all the voters. 
Let $\theta_{-i}$ denote a profile of the types owned by the voters except $i$. 
Let $R(\theta_{i}) = \{\theta'_{i} = (\succ'_{i}, r'_{i}) \mid r'_{i} \subseteq r_{i}\}$ 
denote the set of {\em reportable types} by voter $i$ with true type $\theta_{i}$, assuming that
each $i$ cannot pretend to be a parent of any voter
of which $i$ is not really a parent.
When $i$ reports $r'_{i}$ as her children, we say 
$i$ {\em invites} $r'_{i}$.  
Let $\theta' = (\theta'_{i})_{i \in \mathcal{N}} \in \times_{i \in \mathcal{N}} R(\theta_{i}) = R(\theta)$ denote a reportable type profile.

Given type profile $\theta'$, 
let $\hat{N}(\theta') \subseteq \mathcal{N}$ denote the set of {\em participating voters},
to whom a path exists from $m$ in $G(r'_{\mathcal{N}}, r_{m})$.
%
Given true type profile $\theta$ (which is not observable), 
a social choice function  
$f$ maps each reported profile $\theta' \in R(\theta)$ 
into an outcome $o \in \mathcal{O}$, 
while $f$ can use $(\succ_{i})_{i \in \hat{N}}$ and $(r_{i})_{i \in \hat{N}}$ as parameters.
When the meaning is clear from the context, we will use slightly different notations such as $f(\theta)$ and $f(\succ)$, for the ease
of understanding.

Strategy-proofness over networks~\cite{li:AAAI:2017} is a refinement of strategy-proofness for diffusion mechanism design. It requires that, for each voter, inviting as many children as possible and reporting her true preference is a dominant strategy.

\begin{definition}[Strategy-Proofness over Networks (SP)~\cite{li:AAAI:2017}]
 \label{def:sp-sn}
 An SCF $f$ is 
 {\em strategy-proof over networks} (or satisfy SP in short)
 if 
 for any $N \subseteq \mathcal{N}$ s.t.\ $|N| = n$,
 any $i \in N$,
 any $\theta_{-i} = ((\succ_{j}, r_{j}))_{j \in N \setminus \{i\}}$,
 any $\theta'_{-i} \in R(\theta_{-i})$,
 any $\theta_{i} = (\succ_{i}, r_{i})$,
 and any $\theta'_{i} \in R(\theta_{i})$,
 $f(\theta_{i}, \theta'_{-i})
 \succsim_{i}
 f(\theta'_{i}, \theta'_{-i})$.
\end{definition}

Since one of the main objectives in our paper is to analyze the effect of strategic information diffusion in facility location games,
we also formally define {\em strategy-proofness over networks on information diffusion}, SP-D in short.
This weaker property requires that for any voter, 
who is assumed to report her preference truthfully, 
inviting as many colleagues as possible is a
dominant strategy. 
In practice, such an assumption might be reasonable
for some cases related to social choice and facility location. For example, if citizens are expected to prefer
locations closer to their living addresses, it might
be enough for the decision maker to ask citizens to
invite as many colleagues as possible. SP-D is therefore
a reasonable incentive property for such a situation.

\begin{definition}[Strategy-Proofness over Networks on Information Diffusion (SP-D)]
 \label{def:sp-d}
 An SCF is 
 {\em strategy-proof over networks on information diffusion} (or SP-D in short) 
 if 
 for any $N \subseteq \mathcal{N}$ s.t.\ $|N| = n$,
 any $i \in N$,
 any $\theta_{-i} = ((\succ_{j}, r_{j}))_{j \in N \setminus \{i\}}$,
 any $\theta'_{-i} \in R(\theta_{-i})$,
 any $\theta_{i} = (\succ_{i}, r_{i})$,
 and any $\theta'_{i} = (\succ_{i}, r'_{i}) \in R(\theta_{i})$,
 $f(\theta_{i}, \theta'_{-i})
 \succsim_{i}
 f(\theta'_{i}, \theta'_{-i})$.
\end{definition}

In our model of mechanism design via social network, the original anonymity property requires that any permutation of preferences among
all the voters never changes the outcome, which we sometimes call {\em full-anonymity}.
Now we define further three variants of anonymity properties below 
for SCFs,
namely 
anonymity on structures (AN-S),
anonymity on distances (AN-D),
and 
anonymity on structure-distance pairs (AN-SD).
Briefly speaking, 
AN-S requires that any permutation of preferences among those 
who have the same number of children never changes the outcome,
AN-D requires that any permutation of preferences among those 
who are in the same distance from the moderator never changes the outcome,
AN-SD requires that any permutation of preferences among those 
who have the same number of children and are in the same distance from the moderator never changes the outcome.
By definition, AN implies both AN-S and AN-D,
and both AN-S and AN-D imply AN-SD.

\begin{definition}
 \label{def:anon}
 Given social network $G(r'_{\mathcal{N}}, r_{m})$,
 let $N_{S}(k)$ be the set of participating voters
 who have $k$ children, for each $k \in \mathbb{N}_{\geq 0}$.
 Also, given social network $G(r'_{\mathcal{N}}, r_{m})$,
 let $N_{D}(d)$ be the set of participating voters
 who is at distance $d$ from the moderator, for each $d \in \mathbb{N}_{\geq 0}$.
 An SCF $f$ 
 satisfies {\em AN-S} if,
 for any $\theta$,
 any $k \in \mathbb{N}_{\geq 0}$,
 and any $\theta'$ in which any subset of voters in a certain set $N_{S}(k)$ permutes their preferences 
 from $\theta$, 
 $f(\theta) = f(\theta')$ holds.
 An SCF $f$ 
 satisfies {\em AN-D} if,
 for any $\theta$,
 any $d \in \mathbb{N}_{\geq 0}$,
 and any $\theta'$ in which any subset of voters in a certain set $N_{D}(k)$ permutes their preferences
 from $\theta$,  
 $f(\theta) = f(\theta')$ holds.
 An SCF $f$ 
 satisfies {\em AN-SD} if,
 for any $\theta$,
 any pair $(k,d) \in \mathbb{N}^{2}_{\geq 0}$,
 and any $\theta'$ in which any subset of voters in a certain set $N_{S}(k) \cap N_{D}(d)$ permutes their preferences from $\theta$, 
 $f(\theta) = f(\theta')$ holds.
%
\end{definition}

Note that there is a na\"{i}ve idea for achieving 
those anonymity properties; applying any fully-anonymous GMVS
for the direct-children of the moderator.
Indeed, such an SCF satisfies SP and PE as well.
However, it has an obvious drawback that all the other voters have
no effect {\em at all} on the final outcome.

To avoid this, we also define a parameterized class of {\em voter-relevance}
properties based on the distance from the moderator, which enables us to quantify how far an SCF is from a desirable decision process in which
every voter has some chance to change the final outcome~\footnote{This relevance property discusses to what extent voters' actions affect 
the outcome. 
Similar concepts also exists in the literature, 
such as {\em null player}~\cite{Shapley_1988} and {\em player decisiveness}~\cite{DBLP:journals/scw/LaviMN09},
while these concepts do not quantify the level of violation from some desirable property.}.
We say an SCF is {\em $d$-distance voter-relevant} (or satisfies {\em VR-$d$} in short) if,
for any voter $i$ who is at distance $d$ or less from the moderator,
there exists at least one situation (i.e., a fixed profile of the other voters' actions) $\theta'_{-i}$ in which $i$ can change the outcome by her own action,
e.g., two types $\theta'_{i}$ and $\theta''_{i}$ (see the formal definition below).
In such a case we say voter $i$ is relevant.
From the viewpoint of public decision making,
having a larger $d (\leq n)$ is better. While satisfying VR-0 is meaningless since there is no voter with distance zero from the moderator, we define VR-0 as well for completeness of the discussion.

\begin{definition}
 \label{def:dcs}
 An SCF $f$ is 
 {\em $d$-distance voter-relevant} (or satisfies VR-$d$)
 for some $d \in \mathbb{N}_{\geq 0}$ if,
 for any voter $i \in \bigcup_{1 \leq d' \leq d} N_{D}(d')$ and any $\theta_{i}$,
 there exists $\theta'_{-i}$ s.t.\ 
 $f(\theta'_{i}, \theta'_{-i}) \neq f(\theta''_{i}, \theta'_{-i})$
 holds for some $\theta'_{i}, \theta''_{i} \in R(\theta_{i})$.
\end{definition}

\section{Analyzing Anonymity on Structures}
\label{sec:ANS}

From now on we will present our contributions. Each section corresponds to a column in Table~\ref{tbl:summary}.
We skip AN and begin with AN-S, since just requiring AN-S without any voter-relevance is impossible, along with SP and PE.

\begin{theorem}
 \label{thm:imp:ANS-R0}
 There is no SCF that simultaneously satisfies 
 SP, PE, and AN-S, even without any requirement on the voter-relevance property.
\end{theorem}

\begin{figure}[tb]
 \centering
 \begin{tikzpicture}[scale=0.8]
  \tikzstyle{vote}=[draw, circle, thick, black, fill = white, minimum size = 2mm, inner sep=0pt];
  \tikzstyle{param}=[draw, shape = diamond, thick, black, fill = white, minimum size = 2.5mm, inner sep=0pt];
  \node [style = vote] at (6, 1.2) (m) [label = {0: {$m$}}] {};
  \node [style = vote] at (6, 0.4) (i) [label = {0: {$i$}}] {};
  \node [style = vote] at (6, -0.4) (j) [label = {0: {$j$}}] {};
  \node [style = vote] at (6, -1.2) (u) [label = {0: {$u$}}] {};
  \draw [->, draw = black] (m) -- (i);
  \draw [->, draw = black] (i) -- (j);
  \draw [->, draw = black] (j) -- (u);
  %
  \draw [thick, draw = black] (-3, 1) -- (3,1);
  \draw [thick, draw = black] (-3, -1) -- (3,-1);
  \node at ([yshift = -10pt] -3, 1) {0};
  \node at ([yshift = -10pt] 3, 1) {1};
  \node at ([yshift = -10pt] -3, -1) {0};
  \node at ([yshift = -10pt] 3, -1) {1};
  %
  \node [style=vote] at (1, 1) [label = {90:\small{$p_{i}$}}] {};
  \node [style=vote] at (-1, 1) [label = {90:\small{$p_{j}$}}] {};
  \node [style=vote] at (2, 1) [label = {90:\small{$p_{u}$}}] {};
  %
  \node [style=vote] at (1, -1) [label = {90:\small{$p_{j}$}}] {};
  \node [style=vote] at (-1, -1) [label = {90:\small{$p_{i}$}}] {};
  \node [style=vote] at (2, -1) [label = {90:\small{$p_{u}$}}] {};
 \end{tikzpicture}
 \caption{An example showing that AN-S is incompatible with combination of SP and PE, explained in the proof of Theorem~\ref{thm:imp:ANS-R0}.
 The right diagram indicates the social network among three voters.
 The left top diagram shows the case where voter $i$
 never invites any child,
 and 
 the left bottom shows the case where $i$ invites both children $u$ and $v$.
 The outcome monotonically gets closer to 
 $i$'s peak $p_{i}$.
}
 \label{fig:imp:ANS}
\end{figure}
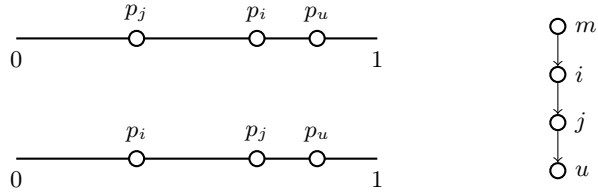

\begin{proof}
 Assume for the sake of contradiction that
 there are three voters $i, j, k$ whose parent-child
 relations are given in the right figure of Fig.~\ref{fig:imp:ANS}.
 We also assume that $p_{i} \neq p_{j}$ holds, e.g., see
 the left top figure of Fig.~\ref{fig:imp:ANS}.

 SP implies SP-D by definition, 
 and SP-D implies, for this example,
 that the facility must be located at the peak of voter $i$;
 otherwise $i$ has an incentive to exclude voter $j$, which also removes $u$.
 This argument is true even when $i$ and $j$ swap their preferences.
 In Fig.~\ref{fig:imp:ANS}, 
 the middle peak is chosen in the left top figure,
 and 
 the leftmost peak is chosen in the left bottom figure.

 Here,
 two voters $i$ and $j$ 
 has the same structure; both have one child. Therefore, AN-S requires that
 swapping their preference does not change the outcome
 when these two voters participate, which violates the above argument.
\qed
\end{proof}

This theorem also has a bit surprising implication; the {\em target rule}, 
which is a well-known GMVS instance and
robust to fake-votes~\cite{todo:AAMAS:2011}, 
also violates the property of SP in our model.
Indeed,
the theorem shows that, when we require AN-S,
any voter-relevance property is not achievable along with SP and PE.
The intuition behind it is that treating
two voters having the same number of children equally,
even if their distances from the moderator are different,
is quite unrealistic from the perspective of diffusion mechanism design. Indeed, to guarantee SP-D in general, 
we should give some priority to those voters closer to the moderator~\footnote{Li et al.~\cite{li:AAAI:2017} handle this issue by proposing a concept of
{\em diffusion critical tree} for the case of auctions.}. Since we further requires AN-S, we end up choosing a fixed alternative as a final outcome, which violates PE.
In the next section we will consider AN-D, a different anonymity concept, and propose a class of SCFs satisfying it along with SP, PE, and VR-2, while AN-S and AN-D have 
no inclusion relation.


\section{Analyzing Anonymity on Distances}
\label{sec:AND}

We now turn to show our analysis on AN-D. As we have mentioned,
AN-D is independently defined from AN-S, and thus the result obtained
in the previous section does not carry over in this section.
Indeed, this section shows a positive result; AN-D is compatible with a certain level of voter-relevance.

We first show a proposition, stating that, when we require SP, PE, and AN-D, 
the outcome must be Pareto efficient even only for 
the direct-children of the moderator. This fact somewhat
corresponds to giving a priority to those voters closer 
to the moderator,
as we have mentioned in the end of the previous section.

\begin{proposition}
 \label{prp:PE:ND1}
 Assume that an SCF $f$ satisfies SP, PE, and AN-D.
 Then, for any input $\theta$, the following holds:
 \[ 
 \min_{i \in N_{D}(1)} p_{i}
 \leq 
 f(\theta)
 \leq
 \max_{j \in N_{D}(1)} p_{j}
 \]
\end{proposition}

\begin{proof}
 We show the statement by mathematical induction on the set of 
 participating voters $\hat{N}$.
 
 As a baseline, let us consider the case where $\hat{N} = N_{D}(1)$,
 i.e., only the set of direct-children of the moderator is participating.
 Let 
 \[
 p^{1}_{\min} := \min_{i \in N_{D}(1)} p_{i}, \ \ \ \ \ 
 p^{1}_{\max} := \max_{j \in N_{D}(1)} p_{j}.
 \]
 From the assumption that $f$ satisfies PE,
 it clearly holds that 
 \[
 p^{1}_{\min} \leq 
 f(\theta)  \leq  p^{1}_{\max}
 \] 
 for arbitrarily chosen input $\theta$ under which only $N_{D}(1)$ is participating.

 We then consider another profile $\theta'$ under which
 another voter $i' \not \in N_{D}(1)$ is 
 participating. 
 Let $i^{*} \in N_{D}(1)$ be the unique ancestor of $i'$. 
 Assume for the sake of contradiction that 
 $f(\theta') < p^{1}_{\min}$.

 Since $f$ satisfies AN-D, the outcome does not change when 
 we change the preferences of voter $i^{*}$ and voter $j$ such that
 $p_{j} := p^{1}_{\max}$. If $i^{*} = j$, i.e., the voter $i^{*}$ 
 is originally having the largest peak among $N_{D}(1)$, we do nothing.

 After this operation, voter $i^{*}$ whose current peak is $p^{1}_{\max}$
 has an incentive to exclude $i'$ (by not inviting $i^{*}$'s child who is an ancestor of $i')$;
 it is guaranteed from the assumption of induction that the outcome is in the range $[p^{1}_{\min}, p^{1}_{\max}]$ when $i^{*}$ exclude $i'$,
 while the outcome $f(\theta')$ under her sincere invitation is 
 strictly less than $p^{1}_{\min}$. This violates the assumption that
 $f$ satisfies SP-D.
 A similar argument works when $p^{1}_{\max} < f(\theta')$.
 \qed
\end{proof}

This proposition might be of an independent interest of readers.
Once we fall into a situation where some voters/agents are specially-treated with a some sort of higher priorities, 
it says that we have to achieve socially-optimal outcome only 
for those agents; otherwise they have incentive to exclude some 
of their collegues from the decision making.

Now we are ready to show our impossibility theorem, which is 
on a combination of an anonymity property AN-D and a voter-relevance 
property VR-2.

\begin{theorem}
 \label{thm:imp:AND-R2}
 There is no SCF that simultaneously satisfies SP, PE,
 AN-D, and VR-2.
\end{theorem}

\begin{proof}
 For the sake of contradiction, assume there exists an SCF $f$
 satisfying SP, PE, AN-D, and VR-2. From VR-2, there exists a
 voter $i \in N_{D}(2)$ such that
 \[
 \exists \theta'_{-i}, \exists \theta'_{i}, \exists \theta''_{i},
 f(\theta'_{i}, \theta'_{-i}) \neq 
 f(\theta''_{i}, \theta'_{-i})
 \]
 holds.
 From the above Proposition~\ref{prp:PE:ND1},
 both $f(\theta'_{i}, \theta'_{-i})$ and
 $f(\theta''_{i}, \theta'_{-i})$ are
 in the range $[p^{1}_{\min}, p^{1}_{\min}]$.

 Since $ f(\theta'_{i}, \theta'_{-i}) \neq 
 f(\theta''_{i}, \theta'_{-i})$ holds,
 at least one of these two outcomes differs
 from $f(\theta'_{-i})$.
 Let us assume without loss of generality that
 \[
 f(\theta'_{i}, \theta'_{-i}) \neq f(\theta'_{-i})
 \]
 Note that, again from Proposition~\ref{prp:PE:ND1},
 $f(\theta'_{-i})$ is also in the range $[p^{1}_{\min}, p^{1}_{\min}]$.

 Now let $i^{*} \in N_{D}(1)$ be the parent of $i$,
 who has the right to exclude $i$.
 If 
 \[
 f(\theta'_{i}, \theta'_{-i}) < f(\theta'_{-i})
 \]
 holds,
 consider swapping preferences of voter $i^{*}$
 and voter $j$ who has the largest peak among $N_{D}(1)$.
 We then apply the same argument with the proof of Proposition~\ref{prp:PE:ND1} and derive a contradiction.
 A similar argument applies for the case
 where $f(\theta'_{i}, \theta'_{-i}) > f(\theta'_{-i})$ holds.
 \qed
\end{proof}

What if we could require a weaker notion of voter-relevance?
The answer is affirmative; if we just require VR-1 as a voter relevance property, we can achieve all the requirement, as the following theorem shows.

\begin{theorem}
 \label{thm:p:AND-R1}
 There is an SCF that satisfies SP, PE,
 AN-D, and VR-1.
\end{theorem}

\begin{proof}
 Consider applying an arbitrarily chosen anonymous generalized median voter scheme by Moulin~\cite{moulin:PC:1980} only for direct-children of the moderator. For example, just choosing the median peak among the reported peaks by the direct-children of the moderator is fine.

 SP trivially holds; for every voter, any invitation strategy
 gives her the same happiness level, which implies the definition of
 SP-D. Furthermore, for the direct-children of the moderator,
 it is well-known that telling their preference truthfully is a best
 strategy in any generalized median voter scheme. All the other voters' preferences have no effect on the outcome. Therefore SP holds.
 
 The outcome is Pareto efficient for $N_{D}(1)$, and thus Pareto 
 efficient for the whole society $\hat{N}$. Therefore the SCF satisfies PE.

 Since we apply an anonymous generalized median voter schemes,
 any preference permutation among $N_{D}(1)$ never changes the outcome.
 Furthermore, it entirely ignore the preferences of all the other voters.
 So any preference permutation among each $N_{D}(d)$ for each $d \geq 2$
 also never changes the outcome. Therefore the SCF satisfies AN-D.

 Finally, by definition of anonymous generalized median voter schemes, any direct-children of the moderator has an instance in which she has a right to choose at least two outcomes. Thus the SCF satisfies VR-1.
 \qed
\end{proof}

Given Theorems~\ref{thm:imp:AND-R2} and \ref{thm:p:AND-R1}, we have found a {\em tight} parameter $d=1$ so that, along with SP, PE, and AN-D, we can achieve VR-$d$ but 
cannot achieve VR-$d+1$.
This is represented in the AN-D column 
of Table~\ref{tbl:summary} in the introduction.
While this is still quite a negative,
there are some sort of flexibility compared to the impossibility on AN-S presented in the previous section.

\section{Analysing Anonymity on Structure-Distance Pairs}
\label{sec:ANSD}

In this section we consider a further weaker variant of anonymity,
namely anonymity on structure-distance pairs (AN-SD),
which is implied by both AN-S and AN-D.
The SCF mentioned in the proof of Theorem~\ref{thm:p:AND-R1}
satisfies AN-SD, but violates VR-2, as any voter who are not directly-connected to the moderator has no effect on the outcome.
Here, we propose an SCF based on an {\em weighted median method},
which satisfies SP, PE, AN-SD, and VR-2.

\begin{theorem}
 \label{thm:p:ANSD-R2}
 There is an SCF that satisfies SP, PE,
 AN-SD, and VR-2. 
\end{theorem}

\begin{figure}[tb]
 \centering
 \begin{tikzpicture}[scale=0.8]
  \tikzstyle{vote}=[draw, circle, thick, black, fill = white, minimum size = 2mm, inner sep=0pt];
  \tikzstyle{param}=[draw, shape = diamond, thick, black, fill = white, minimum size = 2.5mm, inner sep=0pt];
  \node [style = vote] at (6, 1) (m) [label = {90: {$m$}}] {};
  \node [style = vote] at (5, 0) (j) [label = {90: {$j$}}] {};
  \node [style = vote] at (7, 0) (i) [label = {45: {$i$}}] {};
  \node [style = vote] at (6.5, -1) (i1) [label = {135: {$u$}}] {};
  \node [style = vote] at (7.5, -1) (i2) [label = {45: {$v$}}] {};
  \draw [->, draw = black] (m) -- (i);
  \draw [->, draw = black] (m) -- (j);
  \draw [->, draw = black] (i) -- (i1);
  \draw [->, draw = black] (i) -- (i2);
  %
  \draw [thick, draw = black] (-3, 1.5) -- (3,1.5);
  \draw [thick, draw = black] (-3, 0) -- (3,0);
  \draw [thick, draw = black] (-3, -1.5) -- (3,-1.5);
  \node at ([yshift = -10pt] -3, 1.5) {0};
  \node at ([yshift = -10pt] 3, 1.5) {1};
  \node at ([yshift = -10pt] -3, 0) {0};
  \node at ([yshift = -10pt] 3, 0) {1};
  \node at ([yshift = -10pt] -3, -1.5) {0};
  \node at ([yshift = -10pt] 3, -1.5) {1};
  %
  \node [style=vote] at (1, 1.5) [label = {90:\small{$p_{i}$}}, label = {-90: \small{$\times 1$}}] {};
  \node [style=vote] at (-1, 1.5) [label = {90:\small{$p_{j}$}}] {};
  %
  \node [style=vote] at (1, 0) [label = {90:\small{$p_{i}$}}, label = {-90: \small{$\times 2$}}] {};
  \node [style=vote] at (-1, 0) [label = {90:\small{$p_{j}$}}] {};
  \node [style=vote] at (2, 0) [label = {90:\small{$p_{u}$}}] {};
  \node [style=vote] at (1, -1.5) [label = {90:\small{$p_{i}$}}, label = {-90: \small{$\times 3$}}] {};
  \node [style=vote] at (-1, -1.5) [label = {90:\small{$p_{j}$}}] {};
  \node [style=vote] at (2, -1.5) [label = {90:\small{$p_{u}$}}] {};
  \node [style=vote] at (0, -1.5) [label = {90:\small{$p_{v}$}}] {};
 \end{tikzpicture}
 \caption{An example showing that SP-D holds in our proposed weighted median method, explained in the proof of Theorem~\ref{thm:p:ANSD-R2}.
 The right diagram indicates the social network among the four voters.
 The left top diagram shows the case where voter $i$
 never invites any child,
 the left middle shows the case where $i$ only invites a child $u$,
 and 
 the left bottom shows the case where $i$ invites both children $u$ and $v$.
 The outcome monotonically gets closer to 
 $i$'s peak $p_{i}$, which intuitively implies SP-D.
}
 \label{fig:p:WM2}
\end{figure}
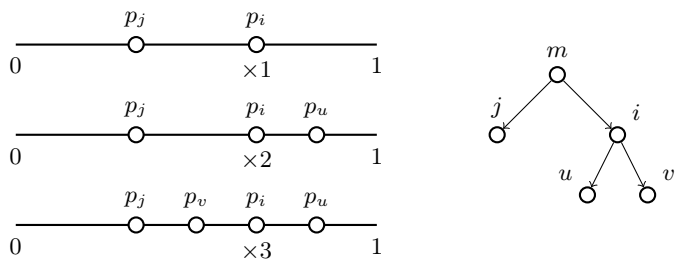

\begin{proof}[Sketch]
 Consider the following SCF $f$; for each voter $i \in \hat{N}$, we first give a weight $w_{i} \in \mathbb{N}_{\geq 1}$ as
 \[
 w_{i} =
 \begin{cases}
  |r'_{i}| + 1 & \text{ if } i \in N_{D}(1) \\
  1 & \text{ if } i \in N_{D}(2) \\
  0 & \text{ otherwise.}
 \end{cases}
 \]
 Then the SCF $f$ takes the median among all the voters peaks, where voter $i$'s peak has $w_{i}$ copies. Formally,
 \[
 f(\theta) = \text{med}(
 \underbrace{p_{1}, \ldots, p_{1}}_{w_{1}},
 \underbrace{p_{2}, \ldots, p_{2}}_{w_{2}},
 \ldots, 
 \underbrace{p_{n}, \ldots, p_{n}}_{w_{n}}),
 \]
 where $p_{i}$ is the peak of voter $i$ who reports preference $\succ_{i}$ and the median operator $\text{med}(\cdots)$ chooses
 the $\lceil m/2 \rceil$-th smallest value among $m$ input values.

 Here we show that the above SCF satisfies all the required properties in the theorem statement.
 PE can be easily shown by the same argument with the proof of Theorem~\ref{thm:p:AND-R1}.

 SP straightforwardly holds for voters in $N_{D}(2)$, with the same argument with Theorem~\ref{thm:p:AND-R1}.
 For those in $N_{D}(1)$, SP-D is guaranteed from the carefully chosen assignment of weights; each voter $i \in N_{D}(1)$ has incentive to invite
 as many children as possible, since having a new child gives her an additional one unit of weight, while the invited child just have one unit of weight. For example, see Fig.~\ref{fig:p:WM2} which explains
 an intuition why SP-D holds. Even in the worst case for voter $i \in N_{D}(1)$,
 where she has a peak at one extreme and all her children are at the other peak, inviting a child never affect the outcome by the definition of the median operator. Moreover,
 her weight is still one unit 
 larger than the sum of the weights of her children. 
 SP is then guaranteed based on the analysis by Moulin~\cite{moulin:PC:1980}, since we can easily show that the above weighted median is an instance of the generalized median voter scheme.

 About AN-SD, observe that any two voters, who are at the same distance from the moderator and invite the same number of children, are assigned the same weight. Therefore, by the definition of $f$, these two voters contribute the outcome in an exactly same way, which guarantees that AN-SD is satisfied.

 We finally show that the SCF satisfies VR-2. It is obvious by definition that any direct-child of the moderator is relevant. For each voter $i$ in $N_{D}(2)$, we can easily find an input in which all direct-children do not invite any of their children except for $i$, and their peaks almost equally distributed between 0 and 1. In such an input, the facility is located at the voter $i$'s peak, which satisfies the condition of relevance; the outcome changes when $i$ reports a different peak location.
 \qed
\end{proof}

Note that the SCF mentioned in the above proof does not satisfy any stronger requirements on both anonymity and voter-relevance. On anonymity,
it violates AN-S from the Theorem~\ref{thm:imp:ANS-R0}. Furthermore,
it also violates AN-D, since the weights can be different among $N_{D}(0)$, according to the number of children. 
On voter-relevance, it violates VR-3, since any report by those who are at distance 3 or more from the moderator is completely ignored.

What if we can totally ignore anonymity properties?
More specifically, to what extent of voter-relevance levels 
can we achieve, besides SP and PE?
As we have already observed,
the SCF mentioned in the proof of Theorem~\ref{thm:p:ANSD-R2} 
satisfies AN-SD and VR-2, which implies that
at least VR-2 is achievable if we totally ignore anonymity.
One of our future directions examines whether there exist SCFs
achieving the full voter-relevance (i.e., VR-n) along with SP and PE.




\section{Discussions and Concluding Remarks}
\label{sec:discussion}

In this paper we focused on deterministic social choice functions 
and discussed to what extent of anonymity and voter-relevance
we can achieve, along with SP and PE.
We provide two impossibility theorems and two existence theorems,
summarized in Table~\ref{tbl:summary}.
We still have open questions in the table; completing them is one of 
our future directions, as mentioned in the previous section.

There are various ways to extend our discussions. For example, what if we can also consider randomized social choice functions? One na\"{i}ve approach in this direction is to randomly choose a dictator, but under this implementation, some voter having many children with completely opposite preferences would remove them to increase the probability that she is chosen as the dictator, which seems to violate SP-D. We believe that, under some 
well-designed probability assignment, we can guarantee the incentive to invite as many children as possible, satisfying SP-D (and also SP) in expectation. 

Extending our discussion to other outcome spaces such as a tree-metric,
a circle metric, a Euclidean space, and/or discrete graphs is also an interesting direction. As many existing works discussed~\cite{schummer:JET:2002,alon:MOR:2010,lu:EC:2010,TodoOY:ECAI:2020}, 
different spaces might have different properties, and the existence of desirable social choice function strictly depends 
on the complexity of the outcome spaces, e.g., the number of vertices/dimensions.
Furthermore, considering weakened incentive properties in facility location games with information diffusion would also be a promising direction, such as non-obvious manipulability~\cite{li:AER:2017,YoshidaKTY:ECAI:2024}.

Last but not least, we strongly believe that both the anonymity properties and the voter-relevance properties defined in this paper are applicable to general diffusion mechanism design framework. The literature of diffusion mechanism design still lacks such a normative analysis. In other words, they have not discussed which decisions are reasonable/applicable and why, and just focused on how to incentivise agents to invite colleagues. As far as we observed, even the reason to increase the market by information diffusion has not been well-justified. Further normative analysis would make the diffusion mechanism design more appealing to the real-life decision making situations in the new era.

\subsection*{Acknowledgments}
\label{ssec:ack}
This work is partially supported by
JST ERATO Grant Number JPMJER2301, 
and
JSPS KAKENHI Grant Numbers JP21H04979 and JP20H00587.


\end{document}